\pdfoutput=1
\documentclass[%
 reprint,
 superscriptaddress,
showpacs,preprintnumbers,
 amsmath,
amssymb,
 aps,
floatfix,
]{revtex4-1}

\usepackage{graphicx}% Include figure files
\usepackage{dcolumn}% Align table columns on decimal point
\usepackage{bm}% bold math
\usepackage{color}
\usepackage{xspace}
\usepackage{lineno}
\usepackage{natbib}
\usepackage{ulem}
\RequirePackage{xspace}

\def\sysp        {\ensuremath{\boldsymbol{f}}}
\def\Kbar  {\kern 0.2em\overline{\kern -0.2em K}{}\xspace}

\def\Kz    {\ensuremath{K^0}\xspace}
\def\Kzb   {\ensuremath{\Kbar^0}\xspace}
\def\KzKzb {\ensuremath{\Kz \kern -0.16em \Kzb}\xspace}
\def\Kp    {\ensuremath{K^+}\xspace}
\def\Km    {\ensuremath{K^-}\xspace}

\def\KpKm  {\ensuremath{\Kp \kern -0.16em \Km}\xspace}

\newcommand{\tev}{\ensuremath{\mathrm{\,Te\kern -0.1em V}}\xspace}
\newcommand{\gev}{\ensuremath{\mathrm{\,Ge\kern -0.1em V}}\xspace}
\newcommand{\mev}{\ensuremath{\mathrm{\,Me\kern -0.1em V}}\xspace}
\newcommand{\kev}{\ensuremath{\mathrm{\,ke\kern -0.1em V}}\xspace}
\newcommand{\ev}{\ensuremath{\mathrm{\,e\kern -0.1em V}}\xspace}
\newcommand{\gevc}{\ensuremath{{\mathrm{\,Ge\kern -0.1em V\!/}c}}\xspace}
\newcommand{\mevc}{\ensuremath{{\mathrm{\,Me\kern -0.1em V\!/}c}}\xspace}
\newcommand{\gevcc}{\ensuremath{{\mathrm{\,Ge\kern -0.1em V\!/}c^2}}\xspace}
\newcommand{\mevcc}{\ensuremath{{\mathrm{\,Me\kern -0.1em V\!/}c^2}}\xspace}
\def\mus  {\ensuremath{\rm \,\mus}\xspace}

\def\gm   {\ensuremath{\rm \,g}\xspace}

\def\mus        {\ensuremath{\,\mu{\rm s}}\xspace}    %% microsecond
\begin{document}

%\linenumbers

\title{Precise Measurement of the Neutrino Mixing Parameter
  $\theta_{23}$ from Muon Neutrino Disappearance in an Off-Axis Beam}

﻿%%%%%%%%%%%%%%%%%%%%%%%%%%%%%%%%%%%%%%%%%%%%%%%%%%%%%%%%%%%%%%
% T2K author list generated on Fri, 28 Feb 2014 07:17:49 +0900
% setting: extra = 0 revtex = 1 yearrule = 1 shiftrule = 1
%         shift rules based on: period1 = 1112 period2 = 1213
% Number of authors = 338
%%%%%%%%%%%%%%%%%%%%%%%%%%%%%%%%%%%%%%%%%%%%%%%%%%%%%%%%%%%%%%

\newcommand{\INSTC}{\affiliation{University of Alberta, Centre for Particle Physics, Department of Physics, Edmonton, Alberta, Canada}}
\newcommand{\INSTEE}{\affiliation{University of Bern, Albert Einstein Center for Fundamental Physics, Laboratory for High Energy Physics (LHEP), Bern, Switzerland}}
\newcommand{\INSTFE}{\affiliation{Boston University, Department of Physics, Boston, Massachusetts, U.S.A.}}
\newcommand{\INSTD}{\affiliation{University of British Columbia, Department of Physics and Astronomy, Vancouver, British Columbia, Canada}}
\newcommand{\INSTGA}{\affiliation{University of California, Irvine, Department of Physics and Astronomy, Irvine, California, U.S.A.}}
\newcommand{\INSTI}{\affiliation{IRFU, CEA Saclay, Gif-sur-Yvette, France}}
\newcommand{\INSTGB}{\affiliation{University of Colorado at Boulder, Department of Physics, Boulder, Colorado, U.S.A.}}
\newcommand{\INSTFG}{\affiliation{Colorado State University, Department of Physics, Fort Collins, Colorado, U.S.A.}}
\newcommand{\INSTFH}{\affiliation{Duke University, Department of Physics, Durham, North Carolina, U.S.A.}}
\newcommand{\INSTBA}{\affiliation{Ecole Polytechnique, IN2P3-CNRS, Laboratoire Leprince-Ringuet, Palaiseau, France }}
\newcommand{\INSTEF}{\affiliation{ETH Zurich, Institute for Particle Physics, Zurich, Switzerland}}
\newcommand{\INSTEG}{\affiliation{University of Geneva, Section de Physique, DPNC, Geneva, Switzerland}}
\newcommand{\INSTDG}{\affiliation{H. Niewodniczanski Institute of Nuclear Physics PAN, Cracow, Poland}}
\newcommand{\INSTCB}{\affiliation{High Energy Accelerator Research Organization (KEK), Tsukuba, Ibaraki, Japan}}
\newcommand{\INSTED}{\affiliation{Institut de Fisica d'Altes Energies (IFAE), Bellaterra (Barcelona), Spain}}
\newcommand{\INSTEC}{\affiliation{IFIC (CSIC \& University of Valencia), Valencia, Spain}}
\newcommand{\INSTEI}{\affiliation{Imperial College London, Department of Physics, London, United Kingdom}}
\newcommand{\INSTGF}{\affiliation{INFN Sezione di Bari and Universit\`a e Politecnico di Bari, Dipartimento Interuniversitario di Fisica, Bari, Italy}}
\newcommand{\INSTBE}{\affiliation{INFN Sezione di Napoli and Universit\`a di Napoli, Dipartimento di Fisica, Napoli, Italy}}
\newcommand{\INSTBF}{\affiliation{INFN Sezione di Padova and Universit\`a di Padova, Dipartimento di Fisica, Padova, Italy}}
\newcommand{\INSTBD}{\affiliation{INFN Sezione di Roma and Universit\`a di Roma ``La Sapienza'', Roma, Italy}}
\newcommand{\INSTEB}{\affiliation{Institute for Nuclear Research of the Russian Academy of Sciences, Moscow, Russia}}
\newcommand{\INSTHA}{\affiliation{Kavli Institute for the Physics and Mathematics of the Universe (WPI), Todai Institutes for Advanced Study, University of Tokyo, Kashiwa, Chiba, Japan}}
\newcommand{\INSTCC}{\affiliation{Kobe University, Kobe, Japan}}
\newcommand{\INSTCD}{\affiliation{Kyoto University, Department of Physics, Kyoto, Japan}}
\newcommand{\INSTEJ}{\affiliation{Lancaster University, Physics Department, Lancaster, United Kingdom}}
\newcommand{\INSTFC}{\affiliation{University of Liverpool, Department of Physics, Liverpool, United Kingdom}}
\newcommand{\INSTFI}{\affiliation{Louisiana State University, Department of Physics and Astronomy, Baton Rouge, Louisiana, U.S.A.}}
\newcommand{\INSTJ}{\affiliation{Universit\'e de Lyon, Universit\'e Claude Bernard Lyon 1, IPN Lyon (IN2P3), Villeurbanne, France}}
\newcommand{\INSTCE}{\affiliation{Miyagi University of Education, Department of Physics, Sendai, Japan}}
\newcommand{\INSTDF}{\affiliation{National Centre for Nuclear Research, Warsaw, Poland}}
\newcommand{\INSTFJ}{\affiliation{State University of New York at Stony Brook, Department of Physics and Astronomy, Stony Brook, New York, U.S.A.}}
\newcommand{\INSTGJ}{\affiliation{Okayama University, Department of Physics, Okayama, Japan}}
\newcommand{\INSTCF}{\affiliation{Osaka City University, Department of Physics, Osaka, Japan}}
\newcommand{\INSTGG}{\affiliation{Oxford University, Department of Physics, Oxford, United Kingdom}}
\newcommand{\INSTBB}{\affiliation{UPMC, Universit\'e Paris Diderot, CNRS/IN2P3, Laboratoire de Physique Nucl\'eaire et de Hautes Energies (LPNHE), Paris, France}}
\newcommand{\INSTGC}{\affiliation{University of Pittsburgh, Department of Physics and Astronomy, Pittsburgh, Pennsylvania, U.S.A.}}
\newcommand{\INSTFA}{\affiliation{Queen Mary University of London, School of Physics and Astronomy, London, United Kingdom}}
\newcommand{\INSTE}{\affiliation{University of Regina, Department of Physics, Regina, Saskatchewan, Canada}}
\newcommand{\INSTGD}{\affiliation{University of Rochester, Department of Physics and Astronomy, Rochester, New York, U.S.A.}}
\newcommand{\INSTBC}{\affiliation{RWTH Aachen University, III. Physikalisches Institut, Aachen, Germany}}
\newcommand{\INSTFB}{\affiliation{University of Sheffield, Department of Physics and Astronomy, Sheffield, United Kingdom}}
\newcommand{\INSTDI}{\affiliation{University of Silesia, Institute of Physics, Katowice, Poland}}
\newcommand{\INSTEH}{\affiliation{STFC, Rutherford Appleton Laboratory, Harwell Oxford,  and  Daresbury Laboratory, Warrington, United Kingdom}}
\newcommand{\INSTCH}{\affiliation{University of Tokyo, Department of Physics, Tokyo, Japan}}
\newcommand{\INSTBJ}{\affiliation{University of Tokyo, Institute for Cosmic Ray Research, Kamioka Observatory, Kamioka, Japan}}
\newcommand{\INSTCG}{\affiliation{University of Tokyo, Institute for Cosmic Ray Research, Research Center for Cosmic Neutrinos, Kashiwa, Japan}}
\newcommand{\INSTGI}{\affiliation{Tokyo Metropolitan University, Department of Physics, Tokyo, Japan}}
\newcommand{\INSTF}{\affiliation{University of Toronto, Department of Physics, Toronto, Ontario, Canada}}
\newcommand{\INSTB}{\affiliation{TRIUMF, Vancouver, British Columbia, Canada}}
\newcommand{\INSTG}{\affiliation{University of Victoria, Department of Physics and Astronomy, Victoria, British Columbia, Canada}}
\newcommand{\INSTDJ}{\affiliation{University of Warsaw, Faculty of Physics, Warsaw, Poland}}
\newcommand{\INSTDH}{\affiliation{Warsaw University of Technology, Institute of Radioelectronics, Warsaw, Poland}}
\newcommand{\INSTFD}{\affiliation{University of Warwick, Department of Physics, Coventry, United Kingdom}}
\newcommand{\INSTGE}{\affiliation{University of Washington, Department of Physics, Seattle, Washington, U.S.A.}}
\newcommand{\INSTGH}{\affiliation{University of Winnipeg, Department of Physics, Winnipeg, Manitoba, Canada}}
\newcommand{\INSTEA}{\affiliation{Wroclaw University, Faculty of Physics and Astronomy, Wroclaw, Poland}}
\newcommand{\INSTH}{\affiliation{York University, Department of Physics and Astronomy, Toronto, Ontario, Canada}}

\INSTC
\INSTEE
\INSTFE
\INSTD
\INSTGA
\INSTI
\INSTGB
\INSTFG
\INSTFH
\INSTBA
\INSTEF
\INSTEG
\INSTDG
\INSTCB
\INSTED
\INSTEC
\INSTEI
\INSTGF
\INSTBE
\INSTBF
\INSTBD
\INSTEB
\INSTHA
\INSTCC
\INSTCD
\INSTEJ
\INSTFC
\INSTFI
\INSTJ
\INSTCE
\INSTDF
\INSTFJ
\INSTGJ
\INSTCF
\INSTGG
\INSTBB
\INSTGC
\INSTFA
\INSTE
\INSTGD
\INSTBC
\INSTFB
\INSTDI
\INSTEH
\INSTCH
\INSTBJ
\INSTCG
\INSTGI
\INSTF
\INSTB
\INSTG
\INSTDJ
\INSTDH
\INSTFD
\INSTGE
\INSTGH
\INSTEA
\INSTH

\author{K.\,Abe}\INSTBJ
\author{J.\,Adam}\INSTFJ
\author{H.\,Aihara}\INSTCH\INSTHA
\author{T.\,Akiri}\INSTFH
\author{C.\,Andreopoulos}\INSTEH
\author{S.\,Aoki}\INSTCC
\author{A.\,Ariga}\INSTEE
\author{T.\,Ariga}\INSTEE
\author{S.\,Assylbekov}\INSTFG
\author{D.\,Autiero}\INSTJ
% ...

\collaboration{The T2K Collaboration}\noaffiliation

\date{\today}% It is always \today, today,

\begin{abstract}
New data from the T2K neutrino oscillation experiment produce the most
precise measurement of the neutrino mixing parameter $\theta_{23}$.
Using an off-axis neutrino beam with a peak energy of 0.6~GeV and a
data set corresponding to $6.57 \times 10^{20}$ protons on target, T2K
has fit the energy-dependent $\nu_\mu$ oscillation probability to
determine oscillation parameters.  The 68\% confidence limit on
$\sin^2(\theta_{23})$ is $0.514^{+0.055}_{-0.056}$ ($0.511 \pm
0.055$), assuming normal (inverted) mass hierarchy.  The best-fit
mass-squared splitting for normal hierarchy is $\Delta m^2_{32} =
(2.51\pm0.10$) $\times 10^{-3}$~eV$^2/c^4$ (inverted hierarchy:
$\Delta m^2_{13} = (2.48\pm0.10$) $\times 10^{-3}$~eV$^2/c^4$).
Adding a model of multinucleon interactions that affect neutrino
energy reconstruction is found to produce only small biases in
neutrino oscillation parameter extraction at current levels of
statistical uncertainty.
\end{abstract}
\maketitle

{\it Introduction.}\textemdash Muon neutrinos 
oscillate to other flavors with a survival probability approximated by
\begin{linenomath}
\begin{align}
P(\nu_\mu \rightarrow \nu_\mu) \simeq& 
    1 - 4 \cos^2(\theta_{13})\sin^2(\theta_{23}) 
  [1-\cos^2(\theta_{13})\nonumber \\
   &\times \sin^2(\theta_{23})] 
\sin^2(1.267\Delta m^2 L/E_\nu),
  \label{eq:oscprob}
\end{align}
\end{linenomath}
where $L(\mathrm{km})$ is the neutrino propagation distance,
$E_\nu(\mathrm{GeV})$ is the neutrino energy and $\Delta m^2
(\mathrm{eV^2/c^4})$ is the relevant neutrino mass-squared splitting:
$\Delta m^2_{32} = m^2_3 - m^2_2$ for normal hierarchy (NH), or
$\Delta m^2_{13} = m_1^2-m_3^2$ for inverted hierarchy (IH).
Oscillation occurs because neutrino flavor eigenstates are linear
superpositions of mass eigenstates, related by a 
mixing matrix parametrized by three mixing angles $\theta_{12}$,
$\theta_{23}$, $\theta_{13}$, and a $CP$ violating phase
$\delta_{CP}$~\cite{Pont1,*Pont2,*Pont3,*MNS}. Previous measurements
\cite{t2knumu_run1-3,PhysRevD.85.031103,PhysRevD.81.092004,Adamson:2012rm,minos-apr2013,MINOS:2013}
have found $\theta_{23} \approx \pi /4$. 
There is considerable interest in precise measurements of
$\theta_{23}$ that can constrain models of neutrino mass
generation (see reviews in
\cite{King:2013eh,Albright:2010ap,Altarelli:2010gt,Ishimori:2010au,Albright:2006cw,Mohapatra:2006gs}), 
determine if $\sin^2 (2\theta_{23})$ is nonmaximal, and, if so whether
$\theta_{23}$ is less or greater than $\pi/4$.

In this Letter, we report the world's most precise measurement of
$\sin^2(\theta_{23})$, using more than twice as much data as our previous
result~\cite{t2knumu_run1-3}, as well as new data selections in T2K's
near detector that measure single pion production processes that can
mimic the oscillation signal in T2K's far detector, Super-Kamiokande (SK).
We also consider the effects of multiple nucleons ejected in
neutrino-nucleus interactions that can cause incorrect neutrino energy
estimates and so affect the oscillation probability
measurement.

{\it T2K experiment.}\textemdash The T2K experiment~\cite{Abe:2011ks}
combines (1) a muon neutrino beam line, (2) near detectors, located
280~m downstream of the neutrino production target, that characterize
the neutrino beam and constrain the neutrino flux parametrization and
cross sections, and (3) the far detector, SK, located
at a distance of $L=295$~km from the target. The neutrino beam axis is
2.5$^\circ$ away from SK, producing a narrow-band
beam~\cite{PhysRevD.87.012001} at the far detector, which reduces
backgrounds from higher-energy neutrino interactions and enhances the
sensitivity to $\theta_{23}$.  The beam's peak energy of
$E_\nu$=2($1.267 \Delta m^2 L/\pi)$ $\approx 0.6$~GeV
corresponds to the first minimum of the $\nu_\mu$ survival probability
at this distance.

A 30 GeV proton beam is extracted in 5~$\mu$s spills from the \mbox{J-PARC} main
ring, directed toward Kamioka in the primary beam line, and hits a
graphite target.  Beam monitors measure the beam's intensity,
trajectory, profile, and beam losses.  Pions and kaons produced in the
target decay into neutrinos in the secondary beam line, which contains
three focusing horns and a 96-m-long decay tunnel. This is followed by
a beam dump and a set of muon monitors.

The near detector complex~\cite{Abe:2011ks} contains an on-axis
Interactive Neutrino Grid detector (INGRID)~\cite{Abe2012} and an
off-axis magnetized detector, ND280. INGRID provides high-statistics
monitoring of the beam intensity, direction, profile, and stability.
ND280 is enclosed in a 0.2 T magnet containing a subdetector optimized
to measure $\pi^0$s (P$\O$D)~\cite{Assylbekov201248}, three time
projection chambers (TPC1,2,3)~\cite{Abgrall:2010hi} alternating with
two one-tonne fine grained detectors (FGD1,2)~\cite{Amaudruz:2012pe},
and an electromagnetic calorimeter~\cite{allan2013electromagnetic} that surrounds the central
detectors. A side muon range detector~\cite{Aoki:2012mf}
identifies muons that exit or stop in the magnet steel.

The SK water-Cherenkov far detector~\cite{Ashie:2005ik} has a 22.5 kt
fiducial volume within a cylindrical inner detector (ID) with 11\,129
inward-facing 20$\,''$ phototubes. Surrounding the ID is a 2 meter
wide outer detector with 1885 outward-facing 8$\,''$ phototubes. A
global positioning system with $<$150 ns precision synchronizes the
timing between SK events and the \mbox{J-PARC} beam spill.

Data were collected during four periods: January-June 2010,
November 2010-March 2011, January-June 2012, and October 2012-May
2013.  The proton beam power on the target steadily increased,
reaching 220 kW with a world record of $1.2 \times 10^{14}$ protons on
target (POT) per spill.  The total neutrino beam exposure on the SK
detector was $6.57 \times 10^{20}$ POT.

 {\it Analysis strategy.}\textemdash The analysis determines oscillation
 parameters by comparing the observed and predicted $\nu_\mu$
 interaction rates and energy spectra at the far detector.  These
 predictions depend on the oscillation parameters, the incident
 neutrino flux, neutrino interaction cross sections, and the detector
 response.

 A measurement of $\nu_\mu$ charged current (CC) events in ND280 is
 used to tune both the initial flux estimates and parameters of the
 neutrino interaction models. The measurement also estimates the
 uncertainties in the predicted neutrino spectrum at the far detector.
 In this new analysis, the ND280 measurement provides better
 constraints on the flux and interaction model parameters by using
 improved event selections, reconstruction, and higher ND280
 statistics. This improvement was achieved by dividing CC events into three
 categories based on the number of pions in the final state.

 At SK, the rate and energy spectrum of $\nu_\mu$ charged current
 quasielastic (CCQE) events are used to determine the oscillation
 parameters through a maximum likelihood fit. The fit accounts for
 uncertainties in the predicted spectrum not only from the
 ND280-constrained flux and interaction models but also SK detector
 selection efficiencies, final state interactions (FSI) inside the
 nucleus, and secondary pion interactions (SI) in the detector
 material.

 {\it Initial neutrino flux model.}\textemdash Detailed simulations of
 hadron production and secondary interactions for primary beam protons
 striking T2K's graphite target predict the neutrino fluxes at ND280
 and SK~\cite{PhysRevD.87.012001}. The simulation is tuned to hadron
 production data such as those from NA61/SHINE for 30~GeV protons on
 graphite~\cite{Abgrall:2011ae, *Abgrall:2011ts}. Pions and kaons
 produced outside the experimentally measured phase space are modeled
 using FLUKA2008~\cite{Ferrari:2005zk, *Battistoni:2007zzb}. The
 GEANT3-based simulations model the horns' magnetic field, particle
 interactions in the horns and decay region, and neutrino production
 from hadron decays.  Flux uncertainties are 10\%-20\%, varying with
 energy, and are dominated by hadron production uncertainties. Full
 covariances between all SK and ND280 energy bins and $\nu$ flavors are
 calculated~\cite{nue2013}.

{\it Neutrino interaction simulations and cross section
Parameters.}\textemdash  
The NEUT Monte Carlo (MC) generator~\cite{Hayato:2009} 
is used to simulate neutrino interactions in T2K's
detectors.  External data, especially from the MiniBooNE
experiment~\cite{mb-ccqe}, set the initial parameters and their
uncertainties subsequently used in the fit to the ND280
data~\cite{nue2013}.  Neutrino interaction parameters fall into two
categories: parameters common between ND280 and SK, and independent
parameters affecting interactions at only one detector.  The common
parameters include the axial masses for CCQE and resonant pion
production, as well as 5 energy-dependent normalizations; these are
included in the fit to the ND280 data, as discussed in the next
section.  Since the ND280 target is mainly carbon while SK's target is
mainly oxygen, additional independent parameters are required which
describe the nuclear model used for CCQE simulation (Fermi momentum,
binding energy and spectral function modeling).  Five additional
cross section parameters relate to pion production, the neutral
current (NC) cross section, the $\nu_e / \nu_\mu$ CC cross section
ratio, and the $\nu / \bar{\nu}$ CC cross section ratio.  These
independent cross section uncertainties (11 parameters) produce a
4.9\% fractional error in the expected number of SK events (see
Table~\ref{tab:nsk_systematic_table_summary}).  Multinucleon 
interactions are thought to lead to an enhancement of the low-energy
cross section
\cite{Marteau:1999kt,Martini:2009,Carlson:2001mp,Shen:2012xz,Bodek:2011}
and have been modeled with a variety of approaches
\cite{Martini:2010,Martini:2013sha,Nieves:2005rq,Benhar:1994hw,Gran:2013kda}.
These effects, not currently included in NEUT, may affect the
 oscillation parameter determination
 \cite{Nieves:2012yz,Lalakulich3,Martini2,Martini3,Meloni,Nieves}. The
 penultimate section presents a study of this potential bias.

{\it ND280 measurements and fits.}\textemdash
The neutrino flux, spectrum, and cross section parameters are
constrained using $\nu_{\mu}$ CC interactions in ND280.  We select the
highest-momentum negatively charged track entering TPC2 with a vertex
inside FGD1's fiducial volume and an energy loss in TPC2
consistent with a muon.  Events originating from interactions in
upstream detectors are vetoed by excluding events with a track in 
TPC1, which is upstream of FGD1.

The ND280 analysis includes many improvements \cite{Abe:2013hdq} over
T2K's previous $\nu_\mu$ disappearance
measurement~\cite{t2knumu_run1-3}, and used a data set with
$5.90\times 10^{20}$ POT.  The selected CC candidate events are now
divided into three samples: \mbox{CC-0$\pi$}, with no identified
pions; CC-1$\pi^+$, with exactly one $\pi^+$ and no $\pi^-$ or
$\pi^0$; and CC-other, with all the other CC events. The
\mbox{CC-0$\pi$} are dominated by CCQE interactions; CC-1$\pi^+$ are
dominated by CC resonant pion production; and CC-other, a mixture of
resonant production and deep inelastic scattering having
final states with $\pi^0$'s, $\pi^-$'s, or multiple pions.  

A $\pi^+$ is identified in one of three ways: an FGD+TPC track with
positive curvature and a TPC charge deposition consistent with a pion,
an FGD-contained track with a charge deposition consistent with a
pion, or a delayed energy deposit due to a decay electron from stopped
$\pi^+ \to \mu^+$ in the FGD.  A $\pi^-$ is tagged only by a
negatively curved FGD+TPC track.  A $\pi^0$ is identified from a track
in the TPC with a charge deposition consistent with an electron from a
$\gamma$ conversion.

The dominant sources of uncertainty are events occurring outside
the FGD fiducial volume and pion reinteractions in the detector.

We fit these three samples for 25 parameters describing the beam flux
in bins of energy and neutrino type at ND280. These parameters
strongly correlate with flux parameters at SK.  We also fit for 19
cross section parameters and for nuisance parameters describing
correlated detector uncertainties of the data bins (10 momentum
$\times$ 7 angle bins for each sample).

\begin{figure}
\includegraphics[width=3.25in]{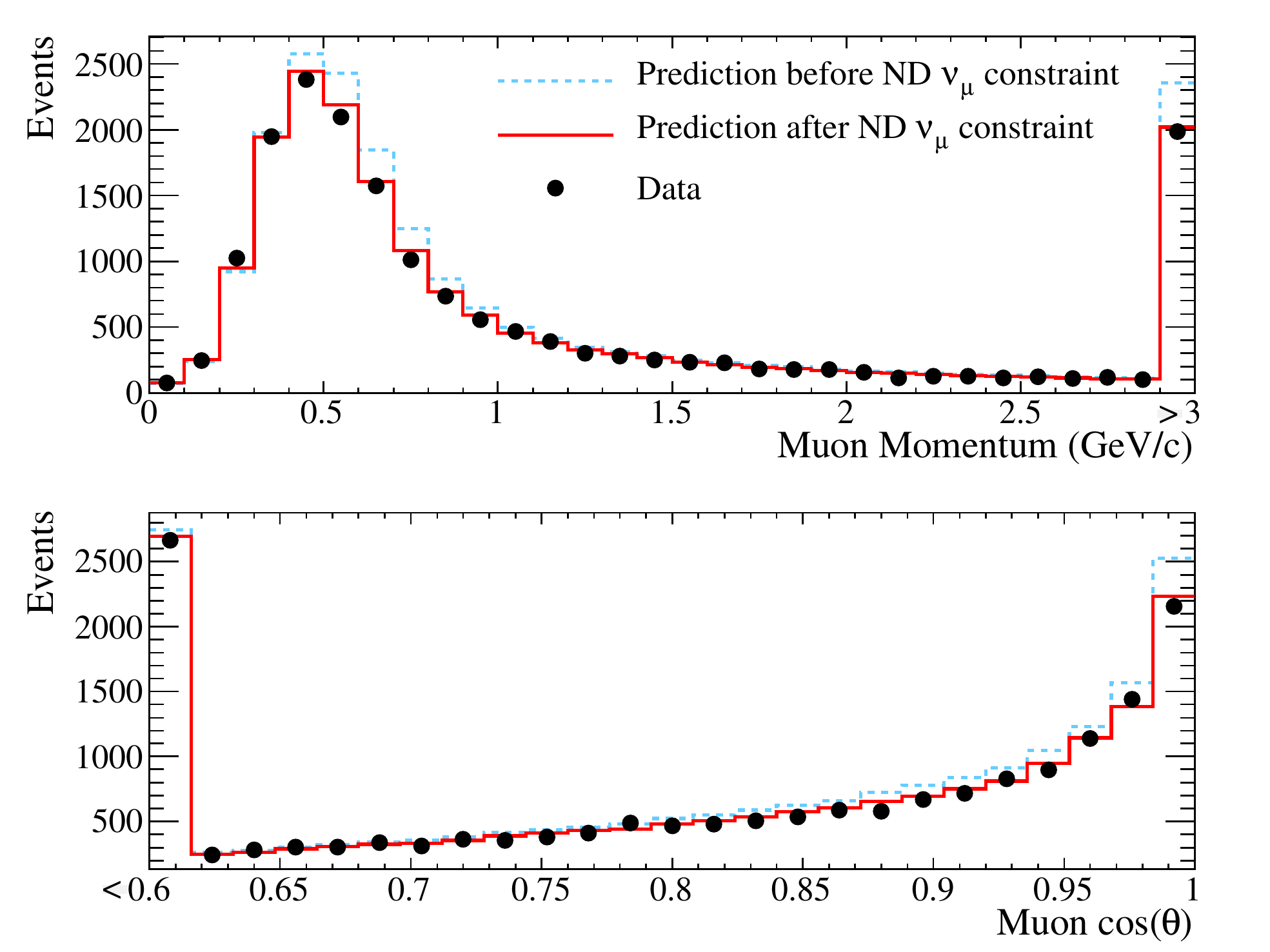}
\caption{\label{fig:nd280pmu-inc} 
The momentum and angular distributions for muons in ND280's \mbox{CC-0$\pi$}
selection.  The predicted distributions before and after the ND280 fit
are overlaid on both figures.
}
\end{figure}

We observe 17\,369, 4047, and 4173 data events in the \mbox{CC-0$\pi$},
CC-1$\pi^+$, and CC-other samples, respectively.  Using the best-fit
parameters, the simulated numbers of events are
17\,352, 4110, and 4119 for the \mbox{CC-0$\pi$}, CC-1$\pi^+$, and CC-other
samples.  Figure~\ref{fig:nd280pmu-inc} shows distributions of the muon
momentum and angle relative to the beam direction
for the \mbox{CC-0$\pi$}
sample and the improvement in data and MC agreement when using the
best-fit parameters. The fit uses the neutrino interaction model to
extrapolate ND280 measurements, primarily covering
$\cos(\theta_\mu)>0.5$, over SK's $4\pi$ angular acceptance.

The fit gives estimates for 16 beam flux parameters at SK, the seven
common cross section parameters, and their covariance. Using the ND280
data reduces the uncertainty on the expected number of events at SK
due to these parameters from 21.6\% to 2.7\%.

 {\it SK measurements.}\textemdash An enriched sample of $\nu_\mu$ CCQE
 events, occurring within -2~$\mu$s to +10~$\mu$s of
 the expected neutrino arrival time, is selected in SK. 
 We require low
 activity in the outer detector to reduce entering backgrounds.  We
 also require: visible energy {$>30$~MeV}, exactly one reconstructed
 Cherenkov ring, $\mu$-like particle identification, reconstructed muon
 momentum {$>200$~MeV}, and $\le 1$ reconstructed decay electron.  The
 reconstructed vertex must be in the fiducial volume (at least 2~m away
 from the ID walls) and we reject ``flasher'' events (produced by
 intermittent light emission inside phototubes).  More details about
 the SK event selection and reconstruction are found in
 \cite{Ashie:2005ik}.

\begin{figure}
  %%\includegraphics[width=8.6cm]{sk_spectrum_and_ratio_per100mev_fine_3pad}
  %%% ratio is 0-6 GeV & has a zoom at 0.25-0.85
  \includegraphics[width=8.6cm]{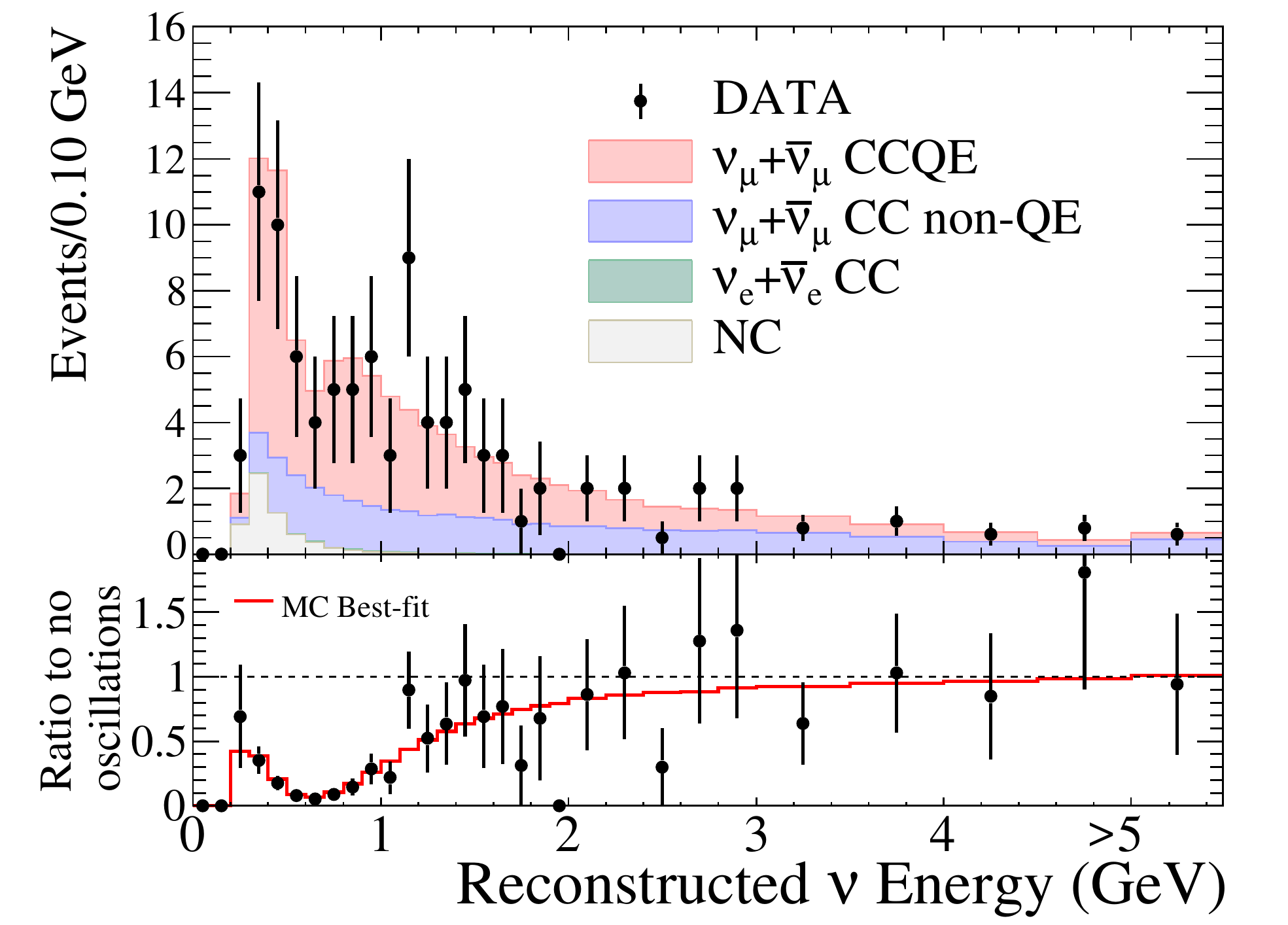} 
  \caption{\label{fig:skspectrum} 
    Reconstructed energy spectrum for single-ring \mbox{$\mu$-like} SK events.
    Top: The observed spectrum and expected spectrum with interaction
    modes for the T2K best fit.
    Bottom: The ratio of the observed spectrum (points) to the
  no-oscillation hypothesis, and the best oscillation fit
    (solid).
  }
\end{figure}

 %Assuming a quasi-elastic interaction with a bound neutron at rest,
 %the neutrino energy is calculated
 %from the reconstructed muon's kinematics using
 %put a blank before eqn to get the lineno's to work
 %\begin{linenomath}
 %\begin{align}
 %E_{\rm reco} = \frac{ m_p^2-(m_n-E_b)^2 - m_\mu^2+ 2(m_n-E_b)E_\mu}{2(m_n-E_b-E_\mu + p_\mu \cos \theta_\mu)},
 %\end{align}
 %\end{linenomath}
 %where $p_\mu$, $E_\mu$, and $\theta_\mu$ are the reconstructed muon
 % momentum, energy, and the angle with respect to the beam direction,
 % respectively; $m_p$, $m_n$, and $m_\mu$ are masses of the proton,
 % neutron, and muon, and $E_b=27$~MeV is the average binding energy of
 % a nucleon in $^{16}$O. 
 The neutrino energy for each event is calculated under the
 quasielastic assumption as in \cite{t2knumu_run1-3} using an average
 binding energy of 27~MeV for nucleons in $^{16}$O.  The $E_{\rm reco}$
 distribution of the 120 selected events is shown in
 Fig.~\ref{fig:skspectrum}.
 The MC expectation without oscillations
  is $446.0 \pm 22.5$ (syst.)
  events, of which 81.0\% are
  $\nu_\mu$+$\bar{\nu}_\mu$ CCQE, 17.5\% are $\nu_\mu$+$\bar{\nu}_\mu$
  CC non-QE, 1.5\% are NC and 0.02\% are $\nu_e$+$\bar{\nu}_e$ CC.  The
  expected resolution in reconstructed energy for
  $\nu_\mu$+$\bar{\nu}_\mu$ CCQE events near the oscillation maximum
  is $\sim$0.1~GeV.

 Systematic uncertainties in the analysis are evaluated with
 atmospheric neutrinos, cosmic-ray muons, and their decay electrons.  
 Correlated selection efficiency parameters are assigned for six event
 categories: $\nu_\mu$+$\bar{\nu}_\mu$ CCQE in three energy bins,
 $\nu_\mu$+$\bar{\nu}_\mu$ CC non-QE, $\nu_e$+$\bar{\nu}_e$ CC, and NC
 events.  An energy scale uncertainty of 2.4\% comes from
 comparing reconstructed momenta in data and MC for
 cosmic-ray stopping muons and associated decay electrons, and
 from comparing reconstructed invariant mass in data and MC
 simulations for $\pi^0$'s
 produced by atmospheric neutrinos.  Systematic uncertainties in pion
 interactions in the target nucleus (FSI) and SK detector (SI) are
 evaluated by varying pion interaction probabilities in the NEUT
 cascade model.
 These SK detector and FSI+SI
 uncertainties produce a 5.6\% fractional error in the expected number
 of SK events (see 
 Table~\ref{tab:nsk_systematic_table_summary}).

%\input{oscfit.tex}
% to be added to symbols.tex
%%%%%%%%%%%%%%%%% Neutrino Mixing   %%%%%%%%%
\def\stst        {\ensuremath{\textrm{sin}^2(2\theta_{12})}\xspace} 
\def\mdmsq     {\ensuremath{|\Delta m^{2}_{32}|\xspace}}
\def\evsqc    {\ensuremath{\rm \,eV^{2} / c^{4}\xspace}}
%%%%%%%%%% Oscillation fit %%%%%%%%%%%%%%%%%%%%%
\def\sysp        {\ensuremath{\boldsymbol{f}}}
\def\nskexp    {\ensuremath{n_{\textrm{SK}}^{\textrm{exp}}}}
\def\nskobs    {\ensuremath{n_{\textrm{SK}}^{\textrm{obs}}}}
\def\nexp    {\ensuremath{n^{\textrm{exp}}}}
\def\nobs    {\ensuremath{n^{\textrm{obs}}}}
\def\LLnd       {\ensuremath{{\cal L}_{\textrm{ND280}}}} 
\def\LLsk       {\ensuremath{{\cal L}_{\textrm{SK}}}} 
%%%%%%%%%%%%%%%%%% LIGHT MESONS  %%%%%%%%%%%%%%%%%
\def\pion  {\ensuremath{\pi}\xspace}
%%%%%%%%%%%% DISTANCE AND AREA %%%%%%%%%%%%%%%%%%%
\def\cmv  {\ensuremath{{\rm \,cm}^3}\xspace}

%%%%%%%%%%%%%%%%%%%%%%%%%%%%%%%%%%%%%%%%%%%%%%%%%%%%%%%%%%%%%%%%%%%%%%%%%%%%%%%%%%%%%%%% 
% best fit spectrum
%%%%%%%%%%%%%%%%%%%%%%%%%%%%%%%%%%%%%%%%%%%%%%%%%%%%%%%%%%%%%%%%%%%%%%%%%%%%%%%%%%%%%%%% 

\begin{table}
  %\begin{ruledtabular}
    \centering
    \begin{tabular}{l | D{.}{.}{-1}}
      \hline \hline
      {\bf Source of uncertainty} (number of parameters)  & \multicolumn{1}{c}{$\delta \nskexp$ / $\nskexp$} \\
      \hline
      ND280-independent cross section (11)              &  4.9\% \\
      Flux and ND280-common cross section (23)           &  2.7\% \\
      SK detector and FSI+SI systematics (7)             &  5.6\% \\
      $\sin^2(\theta_{13})$, $\sin^2(\theta_{12})$, $\Delta m^2_{21}$, $\delta_{CP}$ (4)  &  0.2\% \\
      \hline
      Total (45) & 8.1\%  \\
      \hline \hline
    \end{tabular}
% tmw23 - 1/14/14: Calculated dNsk due to other oscillation parameters.
  %\end{ruledtabular}
  \caption{\label{tab:nsk_systematic_table_summary}
    Effect of 1$\sigma$ systematic parameter variation on
    the number of 1-ring $\mu$-like events,
    computed
    for oscillations with 
    $\sin^2(\theta_{23})=0.500$ and $\mdmsq = 2.40 \times 10^{-3} \evsqc$.
  }
\end{table}

{\it Oscillation fits.}\textemdash 
We estimate oscillation parameters using an unbinned maximum likelihood fit to 
the SK spectrum for the parameters
$\sin^2(\theta_{23})$ and either $\Delta m^2_{32}$ or $\Delta m^2_{13}$ for the normal and inverted mass hierarchies respectively,
and all 45 systematic parameters.
The fit uses 73 unequal-width energy bins, 
%with finer binning near the oscillation peak.
and interpolates the spectrum between bins.
Oscillation probabilities are calculated using the full three-flavor
oscillation framework.
Matter effects are included
with an Earth density of $\rho = 2.6 \gm / \cmv$ \cite{Hagiwara:2011kw},
$\delta_{CP}$ is unconstrained in the range $[-\pi,\pi]$,
and other oscillation parameters are fit with constraints
$\sin^2(\theta_{13}) = 0.0251 \pm 0.0035$,
$\sin^2(\theta_{12}) = 0.312 \pm 0.016$, and 
$\Delta m^2_{21} = (7.50 \pm 0.20) \times 10^{-5}$ eV$^2/c^4$
\cite{pdg2012}. Figure \ref{fig:skspectrum} shows
the best-fit neutrino energy spectrum. 
The point estimates of the 45 nuisance parameters are all within 0.25 standard deviations of their prior values. 

\begin{figure}
\includegraphics[width=8.6cm]{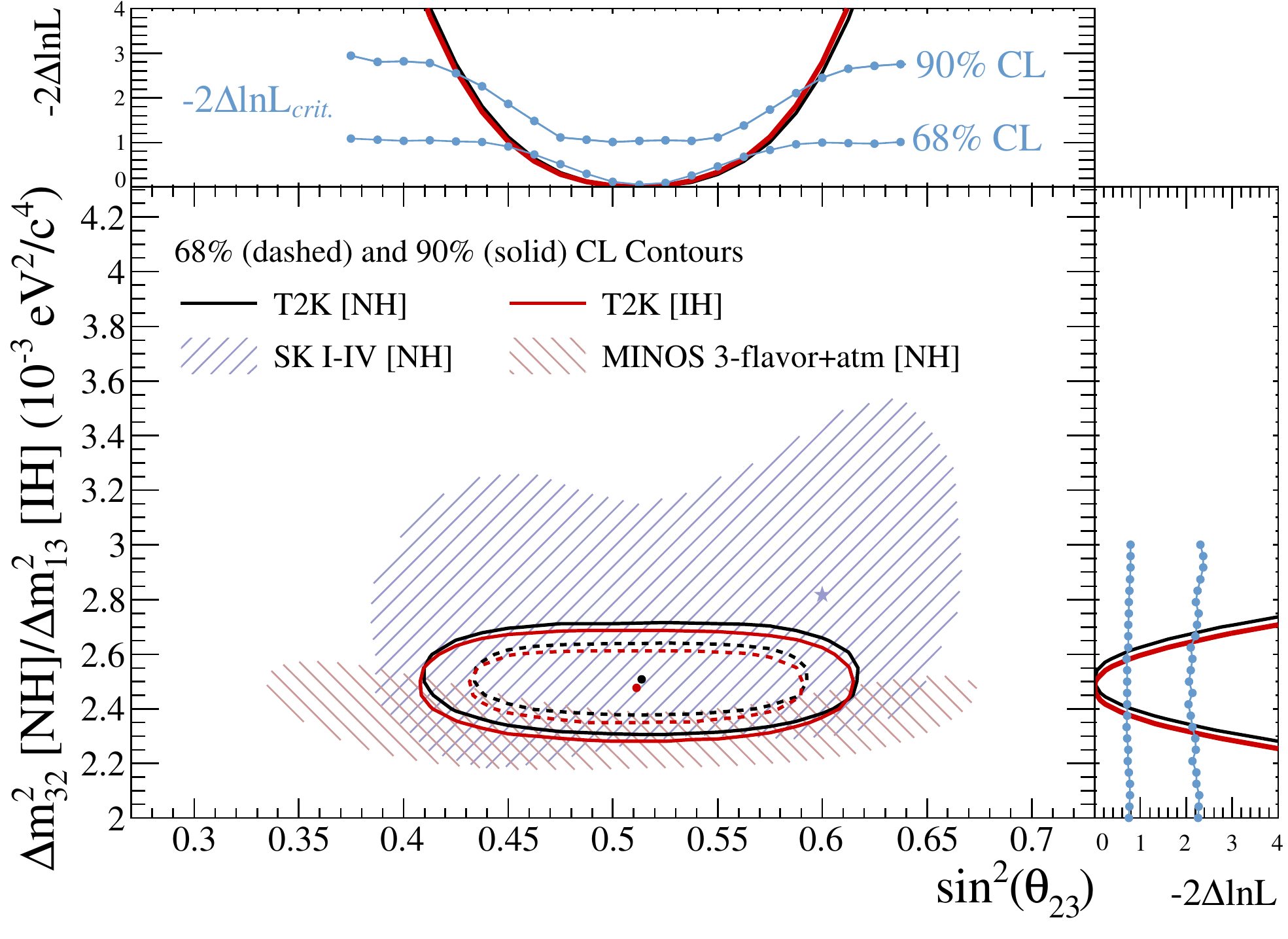} %% T2K+SK+MINOS
  \caption{\label{fig:contour} 
    The 68\% and 90\% C.L. confidence regions for $\sin^2(\theta_{23})$ and $\Delta m^2_{32}$ (NH)
    or $\Delta m^2_{13}$ (IH).
    The SK~\cite{Himmel} 
and MINOS~\cite{MINOS:2013} 90\% C.L. regions
    for NH are shown for comparison.
    T2K's 1D profile likelihoods for each oscillation parameter separately
    are also shown at the top and right overlaid with light blue lines
    and points representing the 1D
    $-2\Delta\ln\mathcal{L}_{\textrm{critical}}$ 
    values for NH at 68\% and 90\% C.L.
}
\end{figure}

Two-dimensional confidence regions in the oscillation parameters are
constructed using the Feldman-Cousins method \cite{PhysRevD.57.3873},
with systematics incorporated using the Cousins-Highland method
\cite{Cousins1992331}.  Figure~\ref{fig:contour} shows
68\% and 90\% confidence regions for the oscillation parameters for
both normal and inverted hierarchies. The 68\% and 90\% expected
sensitivity curves are each 0.04 wider in $\sin^2(\theta_{23})$ than
these contours.  An alternative analysis employing a binned likelihood
ratio gave consistent results. Also shown are 90\% confidence regions
from other recent experimental results.  Statistical uncertainties
dominate T2K's error budget.

We calculate one-dimensional (1D) limits using a new method inspired by
Feldman-Cousins \cite{PhysRevD.57.3873} and Cousins-Highland
\cite{Cousins1992331} that marginalizes over the second oscillation
parameter.  Toy experiments are used to calculate
$-2\Delta\ln\mathcal{L}_{\textrm{critical}}$ values, above
which a parameter value is excluded, for each value of $\sin^2 (\theta_{23})$.  These toy experiments draw values for $\Delta m^2_{32}$
or $\Delta m^2_{13}$ in proportion to the likelihood for fixed 
$\sin^2 (\theta_{23})$, marginalized over systematic parameters.  The toy
experiments draw values of the 45 systematic parameters from either
Gaussian or uniform distributions. We generate $\Delta m^2_{32}$ or
$\Delta m^2_{13}$ limits with the same
procedure. Figure \ref{fig:contour} shows the 1D profile likelihoods for 
both mass hierarchies, with the
$-2\Delta\ln\mathcal{L}_{\textrm{critical}}$ MC estimates for NH.

The 1D 68\% confidence intervals are $\sin^2(\theta_{23}) =
0.514^{+0.055}_{-0.056}$ ($0.511 \pm 0.055$) and $\Delta m^2_{32} =
2.51\pm0.10$ ($\Delta m^2_{13} = 2.48\pm0.10$) $\times 10^{-3} \evsqc$
for the NH (IH).  The best fit corresponds to the maximal possible
disappearance probability for the three-flavor formula.

% % Contour Plot with SK+MINOS

\newcommand{\TK}[1]{\textcolor{red}{#1}}
\hyphenation{multi-nucleon}

\begin{figure}
  \includegraphics[width=8cm]{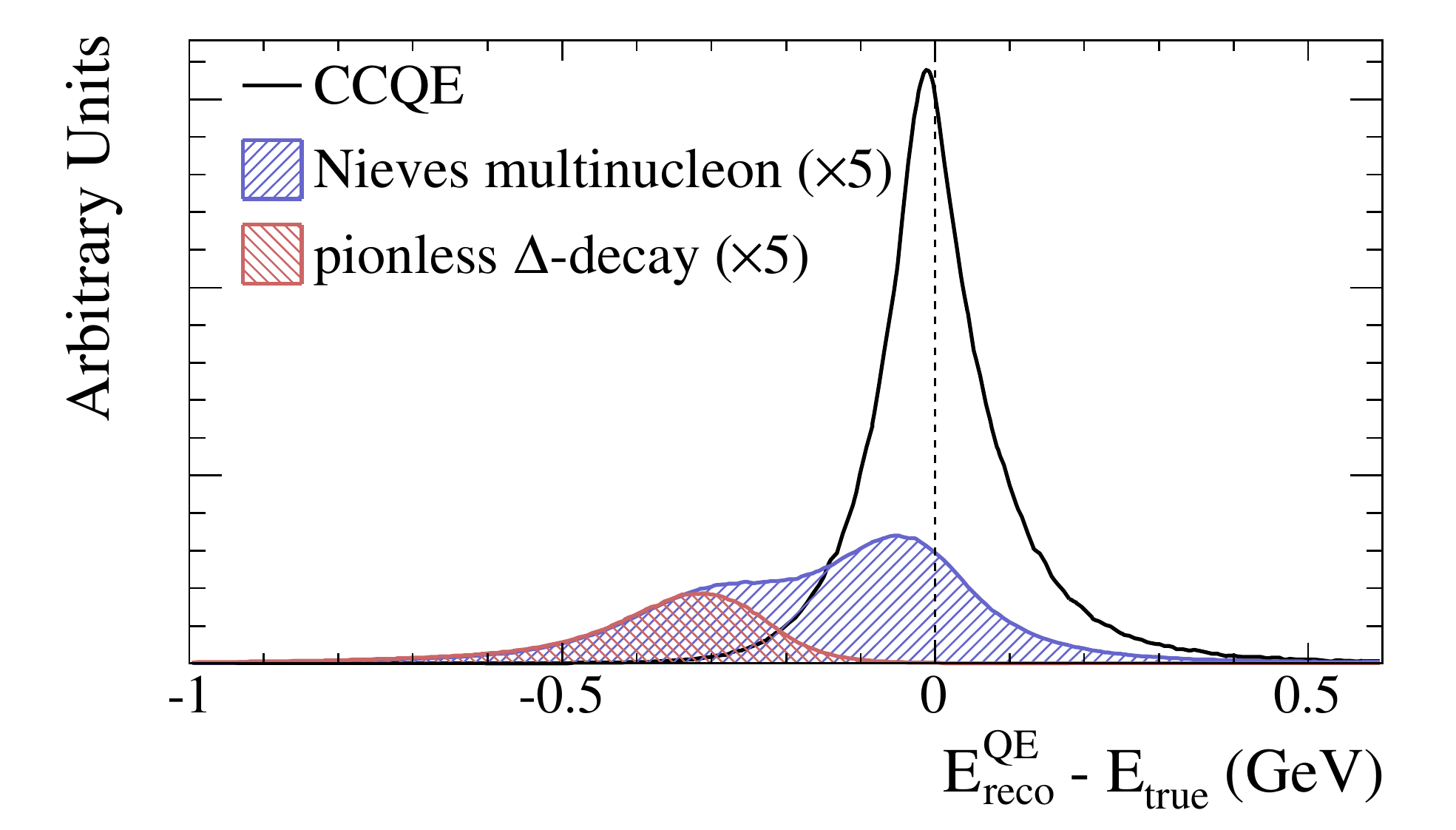}
  \caption{\label{fig:mec} 
	The difference between the reconstructed energy assuming QE
  kinematics and the true neutrino energy. True QE events with
  energies below 1.5~GeV show little bias while multinucleon events based 
  on \cite{Nieves} and NEUT pionless $\Delta$ decay (shown scaled up by a factor
  of 5) are biased towards lower energies.   }
\end{figure}

{\it Effects of multinucleon interactions.}\textemdash
Inspired by more precise measurements of neutrino-nucleus 
scattering~\cite{AguilarArevalo:2007ab, AguilarArevalo:2013hm, 
Fields:2013zhk, Fiorentini:2013ezn}, recent theoretical work suggests 
that neutrino interactions involving multinucleon mechanisms 
may be a significant part of the cross section in
T2K's energy range and might introduce a bias on the oscillation
parameters as large as a few
percent~\cite{Marteau:1999kt, Martini:2009, Carlson:2001mp, Shen:2012xz,
Bodek:2011, Martini:2010, Martini:2013sha, Nieves:2005rq, Benhar:1994hw,
Gran:2013kda, Nieves:2012yz, Lalakulich3, Martini2, Martini3, Meloni, Nieves}.
We are the first oscillation experiment to consider the potential bias
introduced by multinucleon interactions including potential
cancellation from measurements at the near detector. 
At T2K beam energies
most interactions produce final-state nucleons below SK's Cherenkov
threshold, making multinucleon interactions indistinguishable from
quasielastic (QE) interactions.
Even if the additional nucleon does not leave the nucleus, the 
multinucleon mechanism alters the kinematics of the out-going lepton,
distorting the reconstructed neutrino energy which assumes QE kinematics
(see Fig.~\ref{fig:mec}) in addition to increasing the overall QE-like
event rate. 

%At T2K beam energies
%most interactions produce final-state nucleons below SK's Cherenkov
%threshold, making multinucleon interactions indistinguishable from
%quasielastic (QE) interactions.  However,
%the additional nucleon in the interaction 
%alters the kinematics of the out-going lepton, so adding multinucleon
%events to our prediction distorts the reconstructed energy spectrum
%while increasing the overall QE-like event rate. Since T2K
%assumes two-body QE kinematics when reconstructing neutrino
%energies, these events are reconstructed at too low of an
%energy (see Fig.~\ref{fig:mec}).

The T2K neutrino interaction generator, NEUT, includes an effective
model (pionless $\Delta$ decay) that models some but not all of the
expected multinucleon cross section. In order to evaluate the
possible effect on the oscillation analysis, we perform a Monte
Carlo study where the existing effective model is replaced with a
multinucleon prediction based on the work of Nieves~\cite{Nieves}
going up to 1.5~GeV in energy.  
%We used this modified simulation to make ND280 and SK fake data sets
%and performed oscillation analyses as described above on each of
%them, allowing ND280 fake data to renormalize the SK prediction.
%The mean biases in the determined oscillation
%parameters are $<1$\% for this particular multinucleon
%model.
We used this modified simulation to make ND280 and SK fake data sets
with randomly chosen systematic uncertainties but without statistical
fluctuations, and performed oscillation analyses as described above on
each of them, allowing ND280 fake data to renormalize the SK
prediction. The mean biases in the determined oscillation parameters
are $<1$\% for the ensemble, though the $\sin^2(\theta_{23})$ biases
showed a 3.5\% rms spread.

{\it Conclusions.}\textemdash The measurement of
$\sin^2(\theta_{23}) = 0.514^{+0.055}_{-0.056}$ 
($0.511 \pm 0.055$) 
for NH (IH) is consistent with maximal
mixing and is more precise than previous measurements.  
The best-fit
mass-squared splitting is $\Delta m^2_{32} = 2.51\pm0.10$ (IH: $\Delta
m^2_{13} = 2.48\pm0.10$) $\times 10^{-3} \evsqc$.  
Possible multinucleon
knockout in neutrino-nucleus interactions produces a small bias in
the fitted oscillation parameters and is not a significant
uncertainty source at present precision.

We thank the J-PARC staff for superb accelerator performance and the
CERN NA61 collaboration for providing valuable particle production
data. We acknowledge the support of MEXT, Japan; NSERC, NRC, and CFI,
Canada; CEA and CNRS/IN2P3, France; DFG, Germany; INFN, Italy;
National Science Centre (NCN), Poland; RAS, RFBR, and MES,
Russia; MICINN and CPAN, Spain; SNSF and SER, Switzerland; STFC, UK;
and DOE, USA. We also thank CERN for the UA1/NOMAD magnet, DESY for
the HERA-B magnet mover system, NII for SINET4, the WestGrid and
SciNet consortia in Compute Canada, GridPP, UK, and the University of
Oxford Advanced Research Computing (ARC) facility. 
In addition
participation of individual researchers and institutions has been
further supported by funds from: ERC (FP7), EU; JSPS, Japan; Royal
Society, UK; DOE Early Career program, USA.

\bibliographystyle{unsrt}
\bibliographystyle{apsrev4-1}

\bibliography{references}% Produces the bibliography via BibTeX.

\end{document}